\documentclass{article}
\usepackage{arxiv}
\usepackage{natbib}
\usepackage[utf8]{inputenc} 
\usepackage[T1]{fontenc}    
\usepackage{siunitx}

\usepackage{hyperref}       
\usepackage{url}            
\usepackage{booktabs}       
\usepackage{amsfonts}       
\usepackage{nicefrac}       
\usepackage{lipsum}
\usepackage{graphicx}
\usepackage{setspace}
\usepackage{datetime}
\usepackage{fancyhdr}
\usepackage{xcolor}
\usepackage{microtype}

\usepackage{float}
\usepackage{multirow}
\usepackage{subfig}
\usepackage{graphicx}
\usepackage{float}

\graphicspath{ {./images/} }
\usepackage{amsmath}
\usepackage{amssymb,amsmath,amsthm}
\def\tsc#1{\csdef{#1}{\textsc{\lowercase{#1}}\xspace}}
\tsc{WGM}
\tsc{QE}
\tsc{EP}
\tsc{PMS}
\tsc{BEC}
\tsc{DE}
\RequirePackage{colortbl} 
\RequirePackage{soul} 
\newcommand{\highlighting}[1]{\colorbox{yellow}{#1}}

\newcommand{\COM}[1]{}
\def\s{\boldsymbol{s}}

\definecolor{codegreen}{rgb}{0,0.6,0}
\definecolor{codegray}{rgb}{0.5,0.5,0.5}
\definecolor{codepurple}{rgb}{0.58,0,0.82}
\definecolor{backcolour}{rgb}{0.95,0.95,0.92}
\definecolor{c20}{rgb}{0.,0.7,0.}
\definecolor{c30}{rgb}{0.,0.,1.}
\definecolor{c40}{rgb}{1,0.1,0.7}
\definecolor{c50}{rgb}{1,0,0}
\definecolor{c60}{rgb}{1,0.9,0.1}
\definecolor{c70}{rgb}{0.20,0.90,0.30}
\definecolor{c80}{rgb}{0.75,0.5,0.25}
\def\cl#1{\textcolor{c30}{#1}}
\def\cl#1{#1}

\def\CL#1{\textcolor{c50}{#1}}
\def\jpan#1{\textcolor{c20}{#1}}

\def\CL#1{#1}
\def\cling#1{\textcolor{c30}{#1}}
\def\cling#1{#1}
\title{Spatio-temporal Joint Analysis of Extreme Ozone and Moderate PM$_{2.5}$ in California with INLA Approach}

\author{Jianan Pan$^{1,2}$ \quad Kunyang He$^{3}$ \quad Kai Wang$^{3}$\quad Qing Mu$^{4}$\quad Chengxiu Ling$^{3*}$
\\
\small{$^{1}$ School of Mathematics and Physics,   Xi'an Jiaotong-Liverpool University. 111 Ren'ai Road, Suzhou, China, 215123}\\
\small{$^{2}$ Department of Biostatistics,   University of Washington. 1410 Northeast Campus Parkway, Seattle, WA, USA, 98105}  \\
\small{$^{3}$ Academy of Pharmacy,   Xi'an Jiaotong-Liverpool University. 111 Ren'ai Road, Suzhou, China, 215123}\\
\small{$^{4}$ Department of Health and Environmental Sciences, Xi'an Jiaotong-Liverpool University. 111 Ren'ai Road, Suzhou, China, 215123}\\
\small{$^*$ Corresponding author: Chengxiu.Ling@xjtlu.edu.cn}
}

\begin{document}
\maketitle
\begin{abstract}
The substantial threat of concurrent air pollutants to public health is increasingly severe under climate change. To identify the common drivers and extent of spatio-temporal similarity of moderate PM$_{2.5}$ and extreme ozone, this paper proposed a log Gaussian-Gumbel Bayesian hierarchical model allowing for sharing a SPDE-AR(1) spatio-temporal interaction structure. The proposed model outperforms in terms of estimation accuracy and prediction capacity for its increased parsimony and reduced uncertainty, especially for the shared ozone sub-model. Besides the consistently significant influence of temperature  (positive), extreme drought (positive), fire burnt area (positive), and wind speed (negative)  on both PM$_{2.5}$ and ozone,
surface pressure and GDP per capita (precipitation) demonstrate only positive associations with PM$_{2.5}$ (ozone),  while population density relates to neither.  
In addition, our results show the distinct spatio-temporal interactions and different seasonal patterns of PM$_{2.5}$ and ozone, with peaks of PM$_{2.5}$ and ozone in cold and hot seasons, respectively. Finally, with the aid of the excursion function, we see that the areas around the intersection of San Luis Obispo and Santa Barbara counties are likely to exceed the unhealthy ozone level for sensitive groups. Our findings provide new insights for regional and seasonal strategies 
in the co-control of PM$_{2.5}$ and ozone. Our methodology is expected to be utilized when interest lies in multiple interrelated processes in the fields of environment and epidemiology.

\end{abstract}

\section{Introduction} 
Fine particulate matter (PM$_{2.5}$) and ozone (O$_{3}$) are both critical pollutants with severe repercussions 
for global human health \citep{abbafati2020murray}. The simultaneous presence is recognized as a significant public health concern, prompting extensive health-related investigations.
In the United States, elevated ambient ozone levels are associated with increasing mortality rates \citep{hao2015ozone}, while long-term exposure to both PM$_{2.5}$ and ozone pollution is linked to a higher risk of annual hospital admissions \citep{rhee2019impact}. In north India, severe PM$_{2.5}$ and ozone pollution are estimated to cause over 450,000 and 30,000 premature deaths annually, respectively \citep{karambelas2018urban}. 
\cl{Despite rapid reductions (14.97\%) in death
caused by fine particular matter (PM$_{2.5}$) in China from 2015 to 2020, those attributed to ozone increased by 94.61\% during the same period  \citep{guan2021assessing}}.
Furthermore, 
\cl{long-term or short-term concurrent exposure to ozone and PM$_{2.5}$} may lead to more severe health impacts compared to that for individual exposure 
\citep{siddika2019synergistic,gold1999particulate, jerrett2013spatial}.\\
California, with an approximate population of 38,965,000 individuals, stands prominently among states grappbing with severe air pollution, with
six Californian cities in the top ten 
most polluted cities in terms of 
ozone and year-round particle pollution \citep{Lungassociate}. In addition, 
more than half counties (30 and 41 out of 58 counties respectively for ozone and PM$_{2.5}$) in California are classified into counties with the most severe level of air pollution. 
The spatial disparity of air pollution including fine particulate matter, O$_3$, and NO$_2$ and its positive relevance of health issues inn California were identified by \cite{jerrett2013spatial}. We refer to \cite{meo2021effect, mekonnen2021relationship} for the positive association between worse air pollution and increased COVID-19 infection and mortality as well as preterm birth respectively. Additionally, a noticeable correlation has been discovered between both the immediate and prolonged impact of PM$_{2.5}$ concentrations and the rise in mortality rates \cite{shi2016low}, together with a remarkable increase in respiratory emergency department visits due to ozone exposure \cite{malig2016time}. \cl{Given the serious O$_3$ and PM$_{2.5}$ in California, it is essential to identify its generation and spread to implement coordinated prevention and control measures \citep{huang2021strategies, kaufman2020guidance}.}\\
\cl{Bayesian spatio-temporal methods have been} emphasized in air pollution investigations as it can help analyzing \cl{the hotspots of air pollution and incorporate both prior knowledge (e.g., neighbourhood information, the uncertainty of the regression coefficient) and available data}, offering a comprehensive understanding of pollution patterns \citep{craig2008air, gardner2022selecting}. 
\cl{For instance}, PM$_{2.5}$ and ozone pollution episodes in the U.S. are evidenced to be consistent with extreme weather phenomena such as heatwaves and wildfires in space and time \citep{schnell2017co,kalashnikov2022increasing}. \cl{Another advantage of Bayesian methods in the modelling of spatio-temporal data is that they can borrow strength from neighbouring areas to smooth the area-specific risks, and enable the statistical inference for missing data, measurement error, incompatible data and ecological bias} \citep{weber2016assessing, beloconi2018bayesian, zhou2020exploring, blangiardo2013spatial}. However, the complexity of the model and size of the database impose new challenges for using Bayesian methods through the Markov chain Monte Carlo (MCMC) algorithm.  
 To this, the integrated nested Laplace approximation (INLA) approach has been developed as a computationally efficient alternative to MCMC with easy implementation using the R package (\href{https://www.r-inla.org/}{R-INLA package})  \citep{rue2009approximate}. INLA is designed for latent Gaussian fields, including generalized linear mixed, spatial and spatio-temporal models, and have \cl{a great variety of applications} \citep{cameletti2013spatio, koh2023spatiotemporal,chaudhuri2023spatiotemporal,opitz2020point}. 
Furthermore, INLA can be combined with the stochastic partial differential equation (SPDE) approach proposed by \cite{lindgren2011explicit} for incorporating spatial and spatio-temporal structure into \cl{point-reference data, which is widely used for air pollution evaluation \citep{cameletti2013spatio,villejo2023data,cameletti2019bayesian,wang2023spatio,fioravanti2021spatio}} 
\cl{In this study, we will employ Bayesian spatio-temporal methods to jointly analyse moderate PM$_{2.5}$ and extreme ozone level}. 
\CL{The co-pollution of PM$_{2.5}$ and ozone has received increasing attention with emphasis on spatial and temporal patterns. \cite{dai2021co} focused on the correlation of daily average O$_3$-PM$_{2.5}$ in the Yangtze River Delta, China while \cite{schnell2017co, kalashnikov2022increasing} showed extreme cluster of surface ozone, particulate matter, and temperature with consistent offsets in both space and time over eastern North America and U.S., respectively.  However,  there is no study on jointly spatio-temporal modelling for multiple pollutants but only a few research addressed this for PM$_{2.5}$ \citep{wang2023spatio} and ozone \citep{de2018spatial} separately.}  
Note that public health risks and medical burden are more exacerbated by the long-term exposure of PM$_{2.5}$ and severe ozone than the other cases \citep{vicedo2020short}. 
Therefore,  we focus on \cl{monthly average} of  hourly 
PM$_{2.5}$ concentration and monthly maxima of 8-hour {maximum} ozone 
from 181 stations in California from 2017 to 2021, based on the Air Quality Index (AQI) metric reported by the Environmental Protection Agency (EPA) in the United States. 
In addition, we consider to explore the influence of a series of variables on air pollution, namely, 
temperature, precipitation, wind speed, surface pressure, fire burnt area, extreme drought, GDP per capita and population density \citep{kinney2018interactions,jaffe2012ozone}.  
The expected findings of spatio-temporal disparity of air pollution and its influential factors will help identify the hotspots for \cl{comprehensive co-regional risk management} in California with air pollution exceeding the warning risk threshold via excursion function  \citep{bolin2015excursion}, and raise awareness about the control of \cl{multiple} air pollution patterns \cl{in terms of} risk sources, human activities and climate environment. \\
The paper is organized as follows. In Section \ref{section_data}, we \CL{present}  the air quality data  and the predictors. In Section \ref{section_modeling}, we establish the joint spatio-temporal Bayesian hierarchical models for examining the potential spatio-temporal shared pattern between PM$_{2.5}$ and ozone. Model results  are given in Section \ref{sectino_result}. We give an extensive discussion and conclusion  in Section \ref{section_discussion}.

\section{Data Description} \label{section_data}

\subsection{\CL{Air pollution data}} 

California is one of the wealthiest states located on the west coast of the U.S., making up of 58 counties spanning a geographical range with latitudinal and longitudinal coordinates of $32 ^{\circ} 30' N \sim 42 ^{\circ} N$ and $114 ^{\circ}8' W \sim 124 ^{\circ}24' W$.
\begin{figure}[ht]
\centering
\includegraphics[scale=0.15]{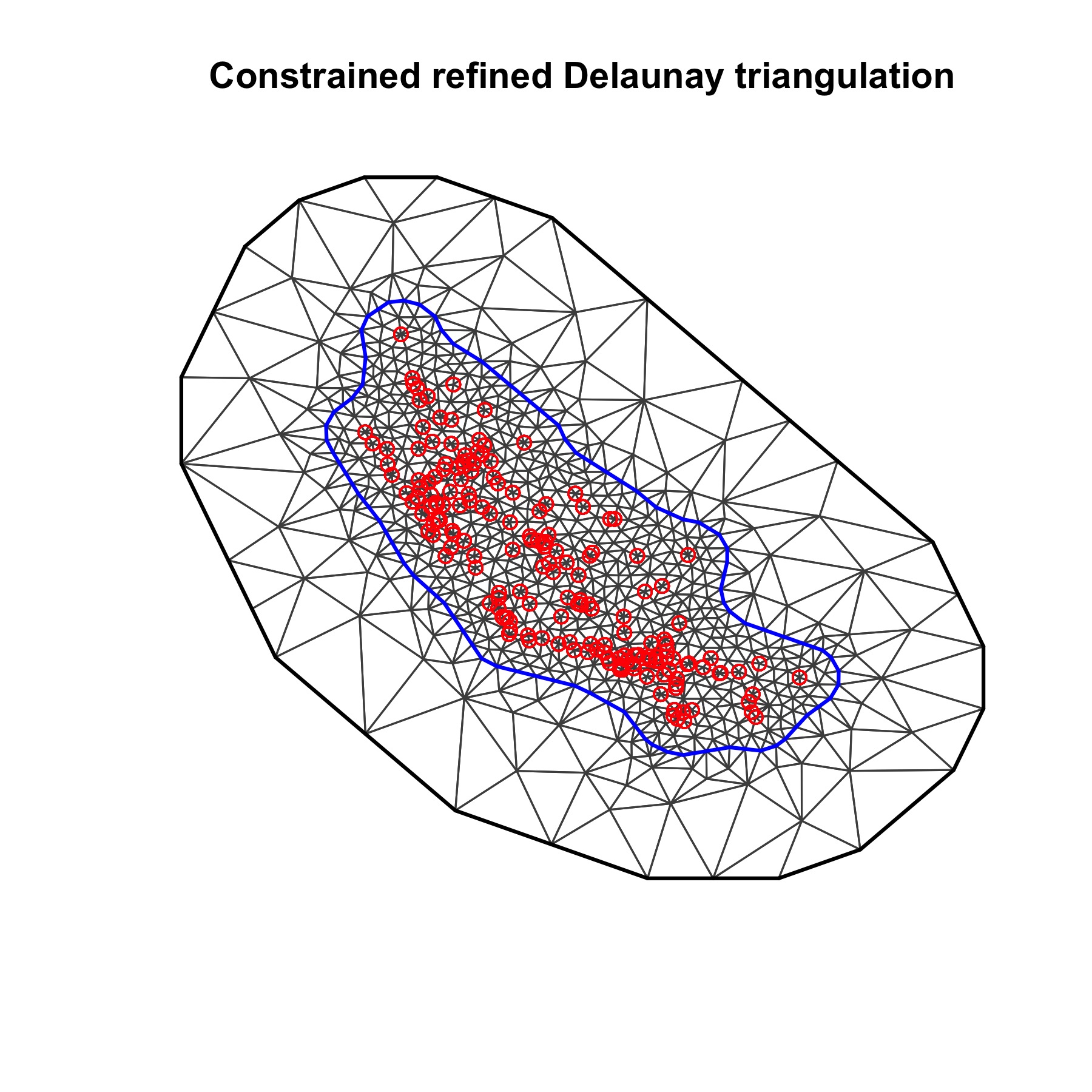}
\caption{Mesh generated for Mat\'ern covariance required in the SPDE approach, with 779 nodes. There are 181 stations in total with 92 stations for PM$_{2.5}$ and 163 stations for ozone.
}
	\label{mesh} 
\end{figure}
We obtained the air pollution dataset including hourly PM$_{2.5}$ levels (with EPA parameter code: 88101, in $\SI{}{\micro\gram/m^3}$) collected by Federal Reference Method (FRM) or Federal Equivalent Method (FEM) observed from 92 weather stations and daily maximum 8-hour ground-level ozone (with EPA parameter code: 44201, in ppb) from 163 stations from 2017 to 2021, accessing \href{https://aqs.epa.gov/qsweb/airdata/download_files.html#Annual}{the United States Environmental Protection Agency (EPA)}. The data is used to generate our response, the monthly mean of PM$_{2.5}$ and the monthly maximum of O$_3$ from the \CL{remaining 181} stations with \cl{more} than 50\% valid data in total. 

\begin{figure*} [htbp]
\centering
	\subfloat[\label{fig2:a}]{
		\includegraphics[scale=0.1]{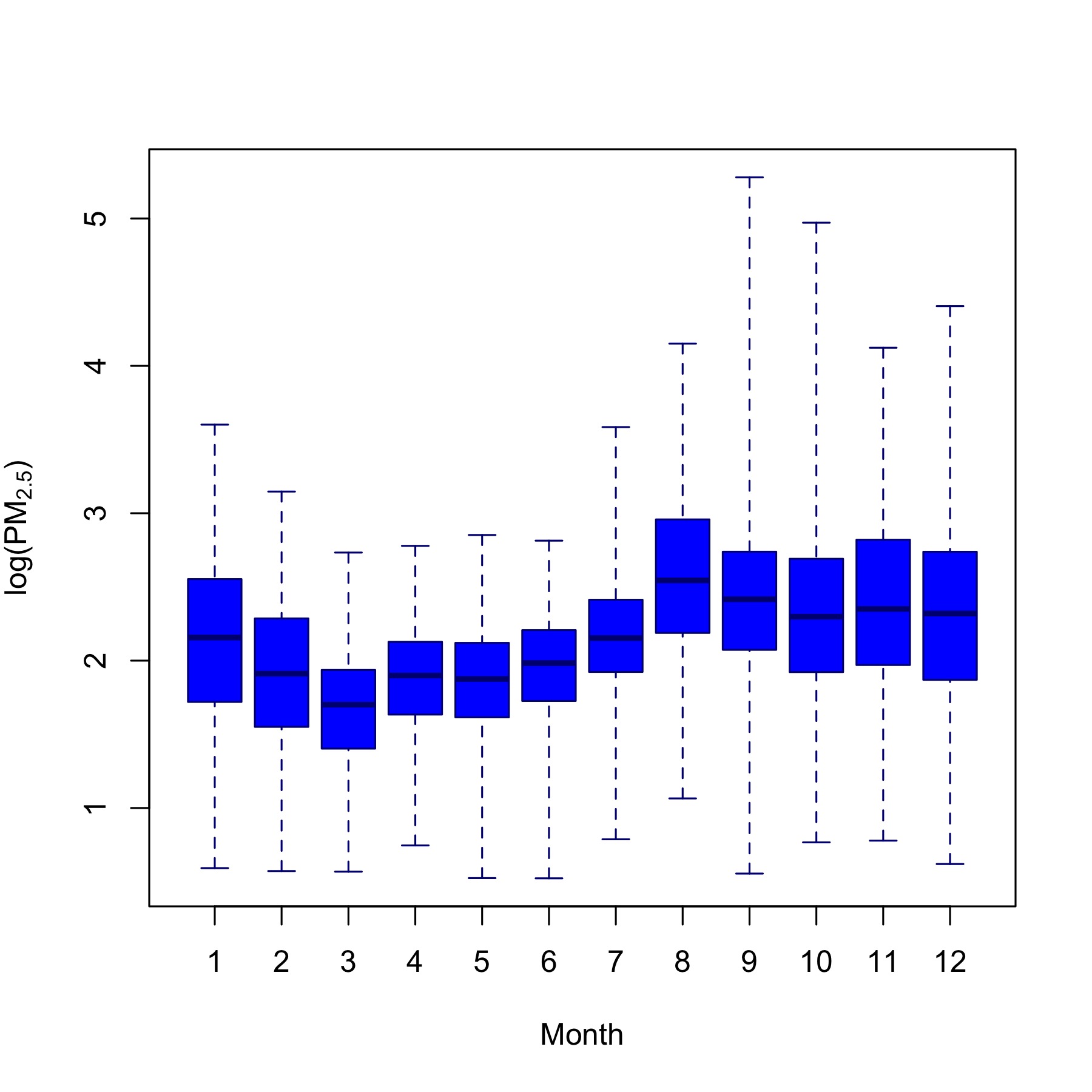}}
	\subfloat[\label{fig2:b}]{
		\includegraphics[scale=0.1]{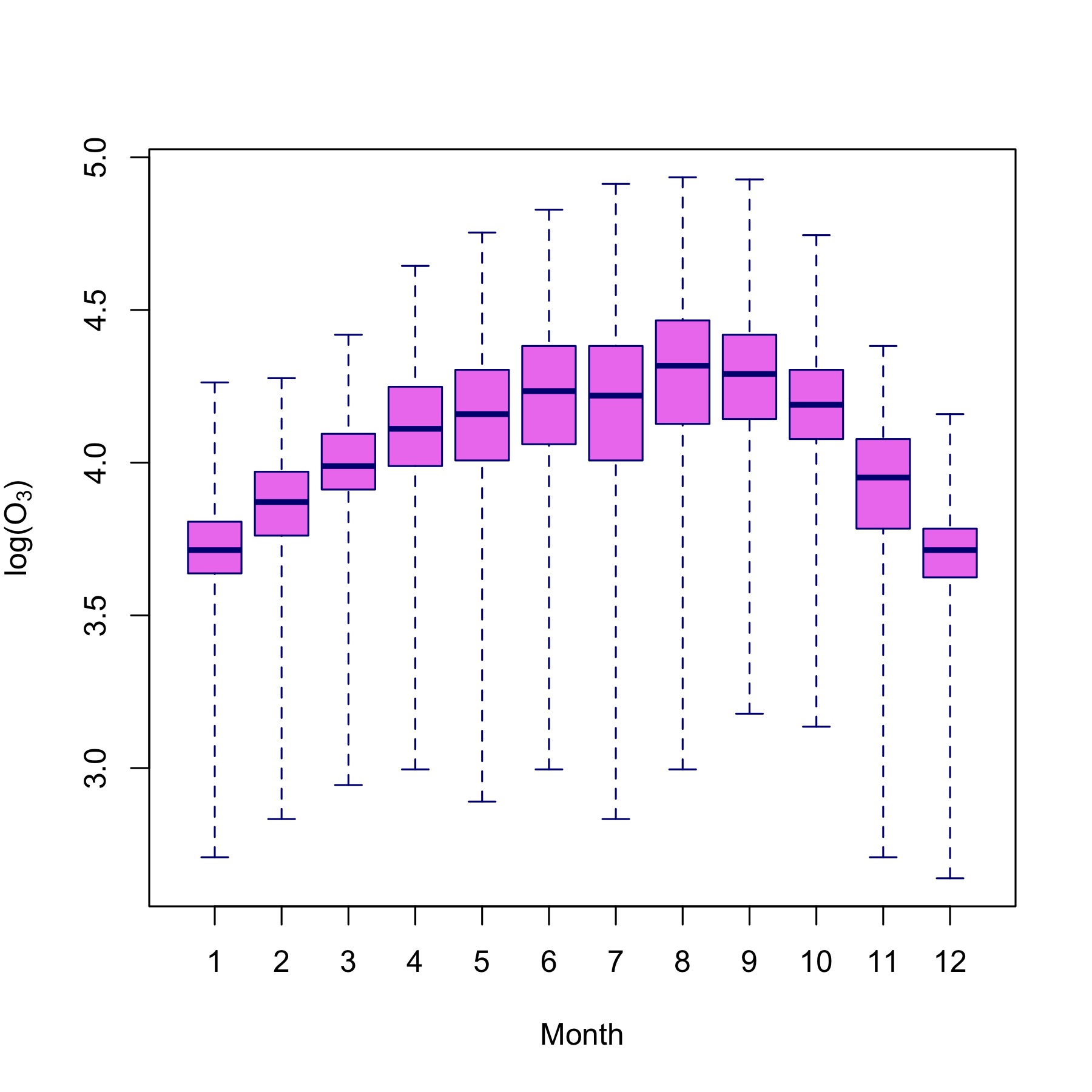}}
\caption{Boxplots of averaged monthly mean PM$_{2.5}$ concentrations (a) and averaged monthly maximum ozone levels (b) on log scale in California during 2017--2021.}
	\label{boxplot} 
\end{figure*}

\begin{figure}[ht]
\centering
        \subfloat[\label{fig1:a}]{
		\includegraphics[scale=0.1]{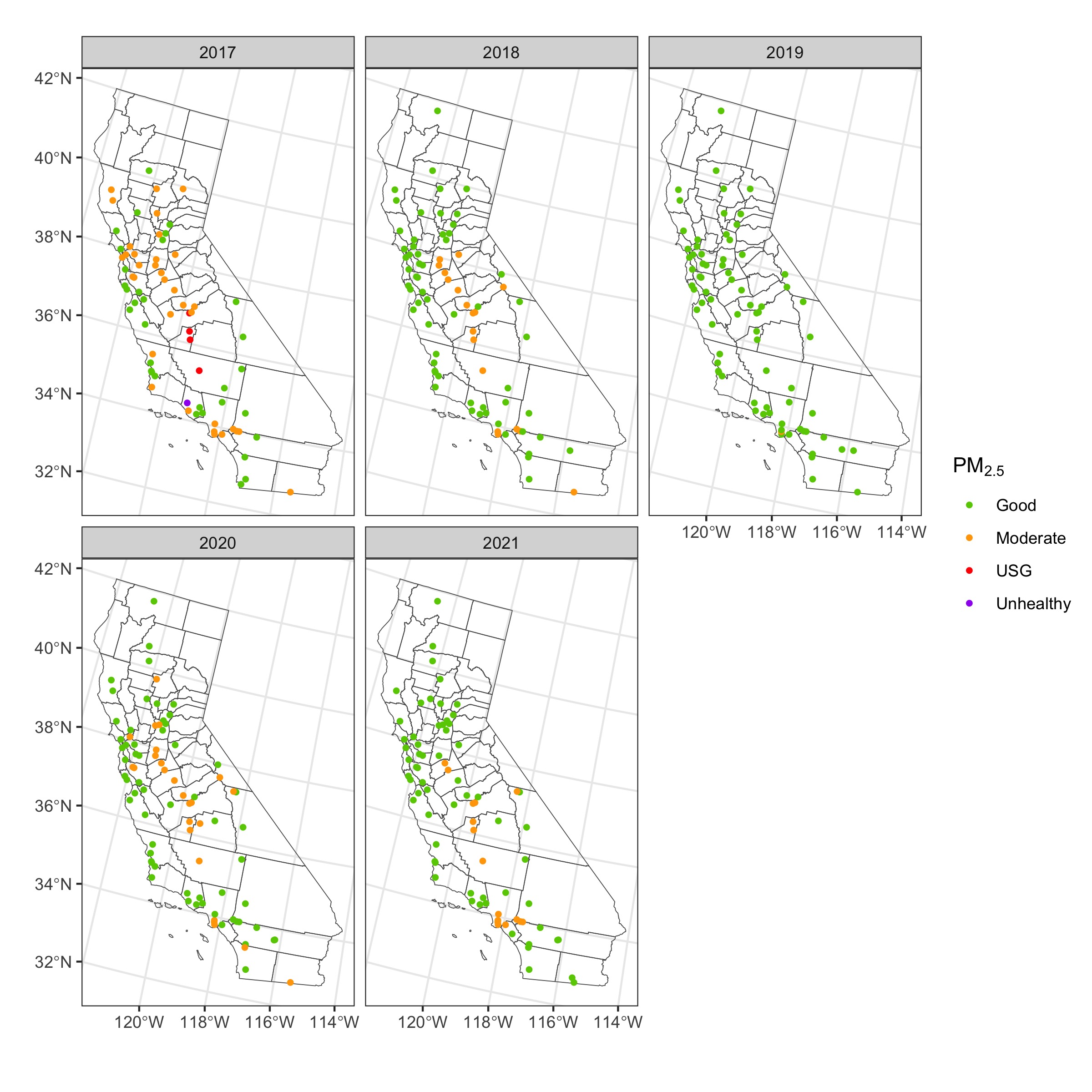}}
	\subfloat[\label{fig1:b}]{
		\includegraphics[scale=0.103]{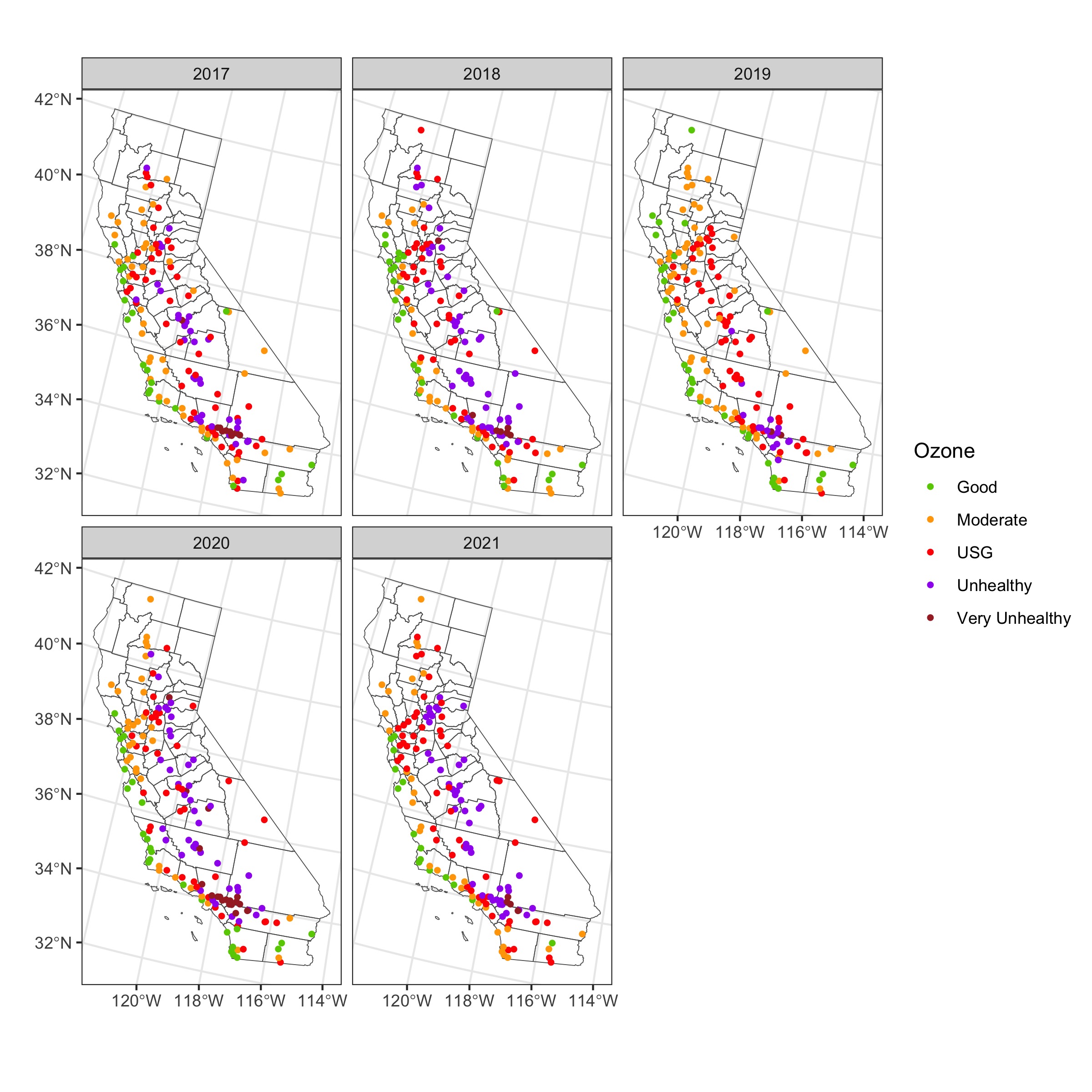}}
\caption{Spatial distribution of  (a) monthly average of daily PM$_{2.5}$ concentrations (in $\SI{}{\micro\gram/m^3}$) in December, \CL{(b)}  monthly max 8-h ozone levels (in ppb) in August in each year during 2017--2021. \CL{Air quality category is given according to the EPA in the U.S.} 
}
	\label{CA_map} 
\end{figure}

\CL{We examined intuitively the temporal and spatial pattern of moderate  PM$_{2.5}$ and extreme ozone in Figs.~\ref{boxplot} and \ref{CA_map}.}  We see that O$_3$ exhibits a conspicuous peak in the summer, while PM$_{2.5}$ seems relatively high during autumn and winter. 
Meanwhile, according to the air quality category of the U.S. EPA, both PM$_{2.5}$ (in \CL{$\si{\micro\gram/m^3}$}) and O$_3$ (in ppb)  can be grouped into \CL{six} levels (PM$_{2.5}$: Good (0--15.4), Moderate (15.5--40.4), Unhealthy for sensitive group (USG; 40.5--65.4), Unhealthy (65.5--150.4), Very Unhealthy (150.5--250.4) and Hazardous (250.5+) and O$_3$: Good (0--54), Moderate (55--70), Unhealthy for Sensitive Groups (USG) (71--85), Unhealthy (86--105), Very Unhealthy (106--200), and Hazardous (201+)). 
From Fig.~\ref{boxplot}, PM$_{2.5}$ concentrations become relatively higher during cold seasons while ozone pollution becomes more serious in summer and autumn.  Fig.~\ref{CA_map} shows the spatial distribution of  PM$_{2.5}$ in December and ozone in August during 2017--2021, California. Except 2019, noticeable PM$_{2.5}$ pollutants in many stations especially in the middle of California are observed at a Moderate or Unhealthy for Sensitive Groups (USG) level, {while over 50\% stations across California experienced serious ozone pollution at a USG or above level.}
These spatial and temporal variations of moderate PM$_{2.5}$ and extreme O$_3$ concentrations {motivate us} to \CL{incorporate spatial and temporal dependence into our Bayesian hierarchical modelling}.

\subsection{Predictors} \label{sec: Predictors}
\cl{Table~\ref{tab: predictors} lists air pollution variables together with 8 considered predictors. It consists of meteorological conditions (surface pressure, precipitation and wind speed), socio-economic factors (GDP per capita, population density), and other factors influencing the generation and spread of the pollutants (fire burnt area, extreme drought).} 

\textit{Meteorological variables}. \cl{Given that} climate and meteorological conditions may affect the generation and spread of air quality \citep{kinney2018interactions},  
we collected {surface pressure (hPa)}, {precipitation (mm)} and {wind speed (10 m/s)} 
 from ERA5-Land dataset \citep{era5land}, which is a reanalysis dataset containing variables on grids with resolution $0.1^{\circ} \times 0.1^{\circ}$ over land surfaces, and  {monthly average daily maximum temperature} from ERA5 global atmospheric reanalysis \citep{era5tmax}. 

\textit{Socioeconomic factors}. To incorporate the socioeconomic factors into account and reflect the influence of human activities, we collected annual GDP per capita (thousands of chained 2012 dollars) and annual {population density (persons per square mile)} in each county in California from 
\href{https://www.bea.gov/data/gdp/gdp-county-metro-and-other-areas}{Bureau of Economic Analysis (BEA), U.S. Department of Commerce}, and  \href{https://www.census.gov/programs-surveys/popest/data/data-sets.2021.List_1725564412.html#list-tab-List_1725564412}{United States Census Bureau}.

\textit{Other factors}. \cl{Noting that} wildfire directly contributes to the generation of PM$_{2.5}$ and ozone \citep{jaffe2012ozone,o2019contribution}, and California is a state experiencing wildfire frequently during the summer and autumn \citep{li2021spatial,o2019contribution}, we included {fire burnt area} representing the severity of the fire from the database provided by \cite{CopernicusClimateChangeService2019}, captured at a grid scale of $0.25^{\circ} \times 0.25^{\circ}$. Drought is also an important aspect of the impact of climate change  on air quality, associated with increased  concentrations of ground-level ozone and PM$_{2.5}$  in the U.S. \citep{wang2017adverse}. We collected the weekly {drought level} from \href{https://droughtmonitor.unl.edu/CurrentMap.aspx}{U.S. Drought Monitor} by the National Drought Mitigation Center (NDMC), the U.S. Department of Agriculture (USDA) and the National Oceanic and Atmospheric Administration (NOA), which categorizes drought severity into six levels: Normal or wet conditions, Abnormally Dry (D0), Moderate Drought (D1), Severe Drought (D2), Extreme Drought (D3) to Exceptionally Drought (D4). \jpan{Each DM class is represented by one polygon saved in provided shape files}. We incorporate an extreme drought indicator  at the monthly scale, with 1 if there is at least one week with extreme drought or above, and 0 otherwise. 

\cl{Noting that the aforementioned covariates are grid-specified (temperature, precipitation, surface pressure, wind speed, fire burnt area) or area specified (\jpan{extreme drought}, GDP per capita and population density), we retrieved these datasets according to the  coordinates of the stations which provided PM$_{2.5}$ and ozone in this study and matched them if  the station's coordinate falls within the grid or area.} In addition, we \CL{confirmed no multi-collinearity} among the covariates included in Table~\ref{tab: predictors} since all  the Variance Inflation Factor (VIF) values  are less than 5 \citep{Salmeron2018}. Note that the substantial dispersion in scales of the continuous predictors listed in Table~\ref{tab: predictors}, we standardize these variables 
to ensure their comparability in the following modelling analysis. 
\begin{table}{}
	\centering
	\caption{Description of variables \CL{with its mean and standard deviation in parenthesis for continuous variables, and frequency for factor variables.}}
	\label{tab: predictors}
        \resizebox{\textwidth}{!}{
	\begin{tabular}{lllll}
		\toprule
		\textbf{Variables  (Unit)}& \textbf{Description } & \textbf{Range} & \textbf{Mean (Std)\CL{/Frequency}} & \textbf{Resolution} \\
		\hline  
  PM$_{2.5} \ (\SI{}{\micro\gram/m^3})$  & Monthly average PM$_{2.5}$ concentration on hourly basis & $0.56 \sim 196.27$ & $10.13 (8.48)$ & station \\
  Ozone (ppb) & Monthly maximum of daily maximum 8-hour ground-level ozone &$14 \sim 139 $ & $ 59.93 (17.38)$ & station \\
	  \multirow{2}{*}{\begin{tabular}{@{}c@{}}Temperature ($\rm{^\circ }C$)
   \end{tabular}} & Monthly average daily maximum temperature &  \multirow{2}{*}{\begin{tabular}{@{}c@{}}$-5.18 \sim 45.09$ 
   \end{tabular}} &
  \multirow{2}{*}{\begin{tabular}{@{}c@{}}$23.16(7.92) $\end{tabular}} &
  \multirow{2}{*}{\begin{tabular}{@{}c@{}}$0.1^\circ \times 0.1^\circ$ \end{tabular}} \\ 
 & of air at 2m above the surface& \\
              Precipitation (mm) & Monthly accumulated precipitation  & $0 \sim 22.67 $ & $ 1.41(2.43)$& $0.1^\circ \times 0.1^\circ$ \\
              Surface pressure (hPa) &  Monthly average pressure of the atmosphere on the surface  &  $730.60 \sim 1027.54 $ & $970.72 (49.24)$& $0.1^\circ \times 0.1^\circ$\\
              Wind speed (10m/s) &  Horizontal wind speed&  $0 \sim 6.02 $& $1.14(0.76)$ & $0.1^\circ \times 0.1^\circ$\\
              Fire burnt area ($\rm{km^2}$) &  Total burnt area within each pixel per month&  $0 \sim 294.86 $ & $0.56(8.21)$ & $0.25^\circ \times 0.25^\circ$\\ 
              \multirow{2}{*}{\begin{tabular}{@{}c@{}} Extreme drought ({dummy variable})\end{tabular}} & Indicator of extreme drought with 1 if at least one weekly  drought&  \multirow{2}{*}{\begin{tabular}{@{}c@{}}0,1\end{tabular}} & No. of 1's: $1573$& \multirow{2}{*}{\begin{tabular}{@{}c@{}}\jpan{polygons in shape files}\end{tabular}}  \\
 & level being D3 or D4 (extreme drought or above), and 0 otherwise  &  & No. of 0's: $9287$\\
              GDP 
              (1000 USD per person) & Annual gross domestic product per person& $27.06 \sim 247.13 $ & $55.39 (27.68)$ & county\\ 
              Population density (person per square miles) & Annual population density& $1.753 \sim 18756.36 $ & $719.24 (1584.68)$ & county\\
	    \bottomrule
    \end{tabular} }
\end{table}

\section{Modelling} \label{section_modeling}


\subsection{Spatio-temporal joint models \label{modeling}}
Let $y_{\text{PM}}(s, t)$ and $y_{\text{OZ}}(s, t)$ denote the logarithmic scales of monthly mean of PM$_{2.5}$ and  monthly maximum of O$_3$ at location $s \in \mathcal{S}$ and time $t \in \mathcal{T}$, where $\mathcal{S}$ represents the study area containing 181 meteorological stations in California and $\mathcal{T} =\{1,2,\ldots, 60\}$ denotes the study period (2017--2021). We establish the following joint spatio-temporal models with mixed effects: 
\begin{equation}
\begin{aligned}\label{Eq: joint}
{\left[y_{\text{PM}}(\boldsymbol{s}, t) \mid \mu_{\text{PM}}(\boldsymbol{s}, t), \sigma_{\text{PM}}\right] } & \sim \operatorname{Gaussian}\left(\mu_{\text{PM}}(\boldsymbol{s}, t), \sigma_{\text{PM}}\right),\\
{\left[y_{\text{OZ}}(\boldsymbol{s}, t) \mid \mu_{\text{OZ}}(\boldsymbol{s}, t), {\sigma_{\text{OZ}}} \right] } & \sim \operatorname{Gumbel}\left(\mu_{\text{OZ}}(\boldsymbol{s}, t), {\sigma_{\text{OZ}}} \right)
\end{aligned}
\end{equation}
with 
\begin{equation}
\label{Eq: fixed}
    \begin{aligned}
       &\mu_{\text{PM}}(\boldsymbol{s}, t) = \boldsymbol{x}(\boldsymbol{s}, t)^{\top} \boldsymbol{\beta}_{\text{PM}}+ f_1(\boldsymbol{s},t) + \epsilon_{1}(\boldsymbol{s},t), \\
&\mu_{\text{OZ}}(\boldsymbol{s}, t) = \boldsymbol{x}(\boldsymbol{s}, t)^{\top} \boldsymbol{\beta}_{\text{OZ}} + f_2(\boldsymbol{s},t) + \epsilon_{2}(\boldsymbol{s},t). 
    \end{aligned}
\end{equation}
As shown in Eq.\eqref{Eq: joint}, we suppose that the logarithmic scales of moderate PM$_{2.5}$ and extreme O$_{3}$ follow respectively Gaussian and Gumbel distributions 
(the generalized extreme value (GEV) distribution with $\xi=0$), respectively \citep{wang2023spatio,blangiardo2013spatial}. The Gumbel distribution is often well-suited for modelling extreme events (block maxima) in environmental studies, such as abnormal precipitation, earthquakes and air quality \citep{rulfova2016two,pisarenko2014characterization,shcherbakov2019forecasting,martins2017extreme}. 

With regard to the spatio-temporal variability of PM$_{2.5}$ and O$_{3}$, we  allow the location parameter ($\mu_{\text{PM}}, \mu_{\text{OZ}}$) in our model (see Eq.\eqref{Eq: fixed}) to vary  according to a combination of fixed effects $(\boldsymbol{\beta}_{\text{PM}},\boldsymbol{\beta}_{\text{OZ}})$, random effects $(f_1, f_2)$, and an independent, \CL{mean zero} Gaussian measurement error $(\epsilon_1, \epsilon_2)$ \CL{with precision $(\tau_1,\tau_2)$}. Here, both fixed effects are associated with the predictor vector $\boldsymbol{x}(\boldsymbol{s}, t)$, consisting of the intercept, temperature, precipitation, surface pressure, wind speed, fire burnt area, extreme drought, GDP per capita and population density. 

In the following, we specify the random effects $(f_1, f_2)$ in Eq.\eqref{Eq: joint} to interpret the spatial and temporal variability in the residuals of  PM$_{2.5}$ and O$_{3}$ after adjusting for the influences of the fixed effects and measurement error. 
\\
\textbf{Model 1. With only fixed effects and measurement errors.} Consider model in Eq.\eqref{Eq: joint} without any other spatial or temporal effects, namely with both $f_1$ and $f_2$ equal zero. In other words, we model PM$_{2.5}$ and O$_{3}$ separately, with only fixed effects associated with predictors and measurement error. We take this as a reference model to compare the model performance in estimation and prediction. 
\\
\textbf{Model 2.} \textbf{With sharing spatial and temporal random effect}. Let  $y(t)$ and $m(t)$  represent the year and month at time $t$, respectively, i.e., $y(t) \in \{1,2,3,4,5\}$ counting the year from 2017 to 2021 and $m(t)\in \{1,2,...,12\}$ counting the month of the year.  
\begin{equation*}
\begin{aligned}
f_1(\boldsymbol{s}, t) & =  u_1(\boldsymbol{s}) + u_2(m(t)) + u_3(y(t)), \\
f_2(\boldsymbol{s}, t) &= \beta_{1}u_1(\boldsymbol{s})+\beta_{2}u_2(m(t)) + \beta_{3}u_3(y(t)) + u_1'(\boldsymbol{s}).
\end{aligned}
\end{equation*}
The spatial random effect ($u_1$) and the temporal random effects ($u_2, u_3$)  are shared between the submodels with scales $\beta_{i}, i=1,2,3$.  The sharing components allow us to model the effects that jointly affect both PM$_{2.5}$ and O$_{3}$ levels. This approach increases model parsimony and reduces the uncertainty in estimating these effects, particularly when observed responses provide insufficient information for complex predictive structures. The scaling factor $\beta_{i}$ also provides some insights into the joint drivers, highlighting the extent of similarities of the spatial and temporal effects in the two processes. Besides the sharing effects, we include an additional spatial term $u_1'$ to capture the extra effect in the O$_3$ sub-model. Both the spatial effects $u_1,u_1'$ are structured by the SPDE approach with a mesh (Fig.~\ref{mesh} constructed with guidance in \cite{Righetto2020}. For example, we suppose that $u_1(\s), \s\in \mathcal S$ a continuous-time Gaussian random field with Mat\'{e}rn covariance structure: For two locations $s_i$ and $s_j$, we have   \citep{cameletti2013spatio}  
\begin{align*} 
\operatorname{Cov}\left(u_1\left(s_{i}\right), u_1\left(s_{j}\right)\right)=
\frac{\sigma^{2}}{2^{\nu-1} \Gamma(\nu)}\left(\sqrt{8 \nu} \frac{h}{\rho}\right)^{\nu} K_{\nu}\left(\sqrt{8 \nu} \frac{h}{\rho}\right)\label{matern},
\end{align*}
where $h$ denotes the Euclidean distance between $s_i$ and $s_j$, $\Gamma$ is the gamma function, $K_{\nu}$ is the modified Bessel function of the
second kind, $\rho > 0$ is the range parameter which denotes the range of non-negligible spatial dependence, $\nu > 0$ is the smoothness parameter, and $\sigma^2 > 0$ is the marginal variance.\\
Meanwhile, the temporal effects $u_2,u_3$ follow the first-order autoregression process (AR(1)), i.e., 
$$u_2(m(t)) =a_m u_2(m(t)-1)+\epsilon^m_t,\quad u_3(y(t)) =a_y u_3(y(t)-1)+\epsilon^y_t$$
with $\epsilon^m_t$ and $\epsilon^y_t$ being Gaussian white noise with variance $\sigma_m^2$ and $\sigma^2_y$, respectively. In addition, we restrict the temporal effect for month ($u_2$) to satisfy the cyclic condition which modified the graph so that the last node ($m(t)=12$) is a neighbour of both $m(t)=11$ and $m(t)=1$.
\\
\textbf{Model 3. With sharing spatio-temporal random effect.} We incorporate the SPDE spatial effect and monthly temporal effects into an SPDE-AR(1) model \citep{cameletti2013spatio}. Compared with Model 2, we combined the SPDE spatial random effect and the monthly temporal effect as follows.
\begin{equation*}
\begin{aligned}
f_1(\boldsymbol{s}, t) & = u_4(y(t)) + u_5(\boldsymbol{s}, m(t)), \\
f_2(\boldsymbol{s}, t) &=  \beta_{4}u_4(y(t)) + \beta_{5}u_5(\boldsymbol{s}, m(t)) + u_5'(\boldsymbol{s}, m(t)). 
\end{aligned}
\end{equation*}
Here $u_4$ is the yearly AR(1) model similar to $u_3$ in Model 2. The spatio-temporal random effect $u_5(\boldsymbol{s}, m(t))$ follows a SPDE-AR(1) model proposed by \cite{cameletti2013spatio}. In particular,
\begin{align*}
u_5(\boldsymbol{s}, m(t))  &=\cl{a_s} u_5(\boldsymbol{s},  m(t) -1)+w(\boldsymbol{s},  m(t)) \label{ar1},\\
w(\boldsymbol{s},m(t))  &\sim \mathcal{G} \mathcal{P}_{2 \mathrm{D}-\operatorname{SPDE}}\left(\rho_{s}, \sigma_{s}, \nu_{s}\right).
\end{align*}
We see that the $u_5$ varies in time via a monthly cyclic AR(1) model over a year with coefficient $a_{\s}$ and its innovation part $w(\boldsymbol{s}, m(t))$ follows a two-dimensional SPDE model \cl{in space and independent at different time points} 
\citep{lindgren2011explicit}. Similar to Model 2, the additional term $u_5'(\boldsymbol{s}, m(t))$ has the same SPDE-AR(1) structure as $u_5$. While the sharing coefficient $\beta_5$ reveals the similar/inverse spatiotemporal effect between the two processes, the additional term $u_5'$ increases both model fitness and credibility of inference.

\subsection{Prior definition} \label{prior} 
A typical property of Bayesian inference is that the prior distribution is required during the calculation of posterior density. We define vague Gaussian priors for the fixed effects $\boldsymbol{\beta}_{\text{PM}},\boldsymbol{\beta}_{\text{OZ}}$, the sharing coefficients $\beta_i, i=1,\ldots,5$, and the variance parameters $\sigma^2_{m}, \sigma^2_{y}$ in the AR(1) temporal model.
Note that INLA often defines the precision parameter with $\tau=1/\sigma^2$, we set log-gamma priors for $\tau_{\text{PM}}$ (in Gaussian likelihood), $\tau_{\text{OZ}}$ (in Gumbel likelihood) as well as $\tau_{1}, \tau_{2}$ in the measurement errors. The smooth parameter in the SPDE approach is commonly fixed with $\nu_{s}=1$ in spatial analysis.  For other parameters, we employ the penalized complexity (PC) priors \citep{simpson2017penalising} that penalize the complex models to avoid overfitness. Following the relevant research \citep{fuglstad2019constructing}, the PC priors for the SPDE and SPDE-AR(1) processes are defined with $\operatorname{Prob}\left(\rho_{s}<10\right)=0.9$, $\operatorname{Prob}\left(\sigma_{s}>0.5\right)=0.1$, which means
the spatial dependence is unignorable within the range of 10km and the probability of the standard deviation larger than 0.5 is very low. For the AR(1) models, we set $\operatorname{Prob}\left( a_s > 0\right)=0.95 \nonumber$ for the SPDE-AR(1) model and \cl{$\operatorname{Prob}\left( a_m> 0\right)= \operatorname{Prob}\left( a_y> 0\right) = 0.95$} for other monthly and yearly AR(1) models. 
The sensitivity analysis utilized default priors in INLA and varied spatial nodes in the SPDE approach. The results confirm the robustness of of all the priors we defined, with consistent posterior estimations of fixed effects and parameters.

\section{Results} \label{sectino_result}



\subsection{Model comparison} \label{model_result}
We carried out the three joint models \cl{proposed in Section \ref{modeling} through the implementation of INLA.} The performance of the model's estimation on the training set (2017--2020) and prediction on the validation set (2021) is compared. Model performance on the training set is assessed using the Watanabe-Akaike information criterion \citep[WAIC;][]{watanabe2010asymptotic}, deviance information criterion \citep[DIC;][]{spiegelhalter2002bayesian}, and the conditional predictive ordinate \citep[CPO;][]{lewis2014posterior} with its summative form of logarithm score \citep[LS;][]{Gneiting2007}. 
For the validation set, we utilize the correlation coefficient to measure the linear correlation between predicted and observed values, and the root mean square error (RMSE) \cl{for the magnitude of its difference}, and introduce scaled threshold weighted continuous ranked probability score \citep[StwCRPS;][]{bolin2023,Vandeskog2022} to evaluate \cl{the model predictive power in the tail parts, namely the model fitness} of extreme O$_{3}$. 
All the criteria except the correlation exhibit preferences for lower values.

The evaluation of the performance is shown in Table~\ref{DIC}. On the training set, Model 3 exhibits superior performance with lower DIC, WAIC and LS. The validation 
\cl{for PM$_{2.5}$ and O$_{3}$ is evaluated separately.} 
All three O$_{3}$ sub-models perform better than the PM$_{2.5}$ models, achieving correlations over 70\% and RMSE below 0.20. This can be attributed to the availability of O$_{3}$ data, enhancing the reliability of O$_{3}$ prediction. Despite Model 3 failing to excel in StwCRPS on the O$_{3}$ validation set, the performance is favourable in both correlation and RMSE, and it is still selected as the optimal model for the subsequent analysis. 

\begin{figure} [htbp] 
\centering
	\subfloat[]{
		\includegraphics[scale=0.1]{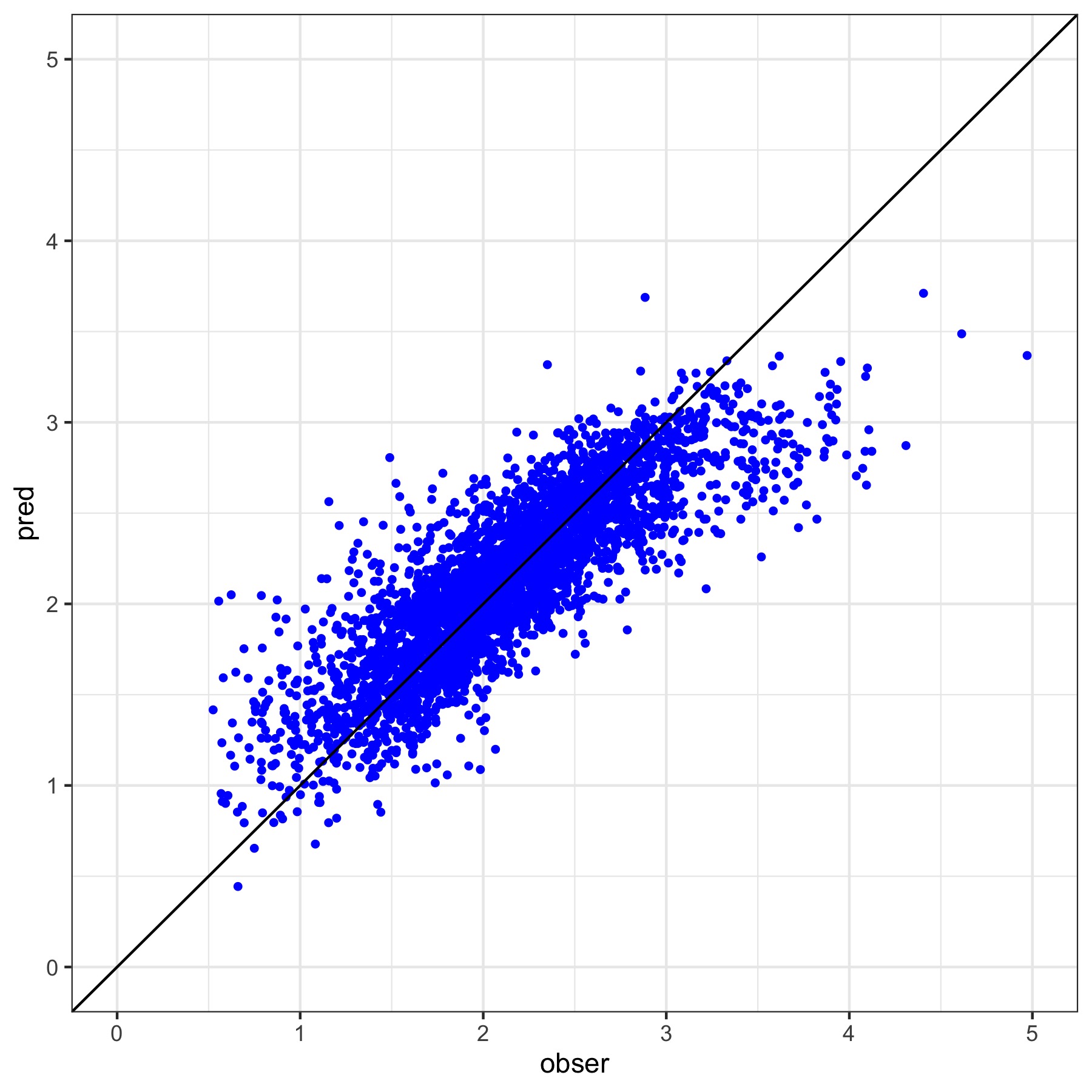}}
	\subfloat[]{
		\includegraphics[scale=0.1]{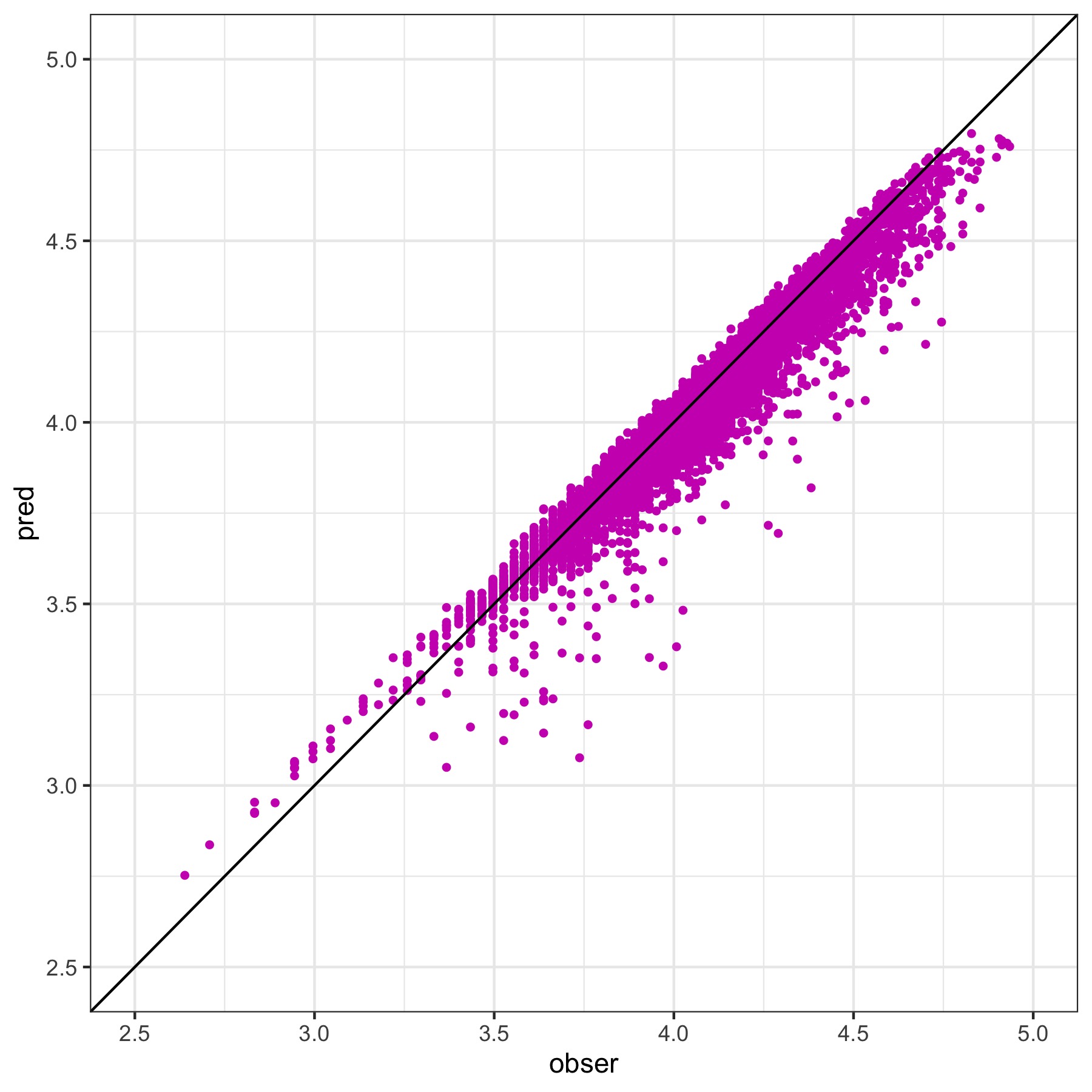}}\\
  \subfloat[]{
		\includegraphics[scale=0.1]{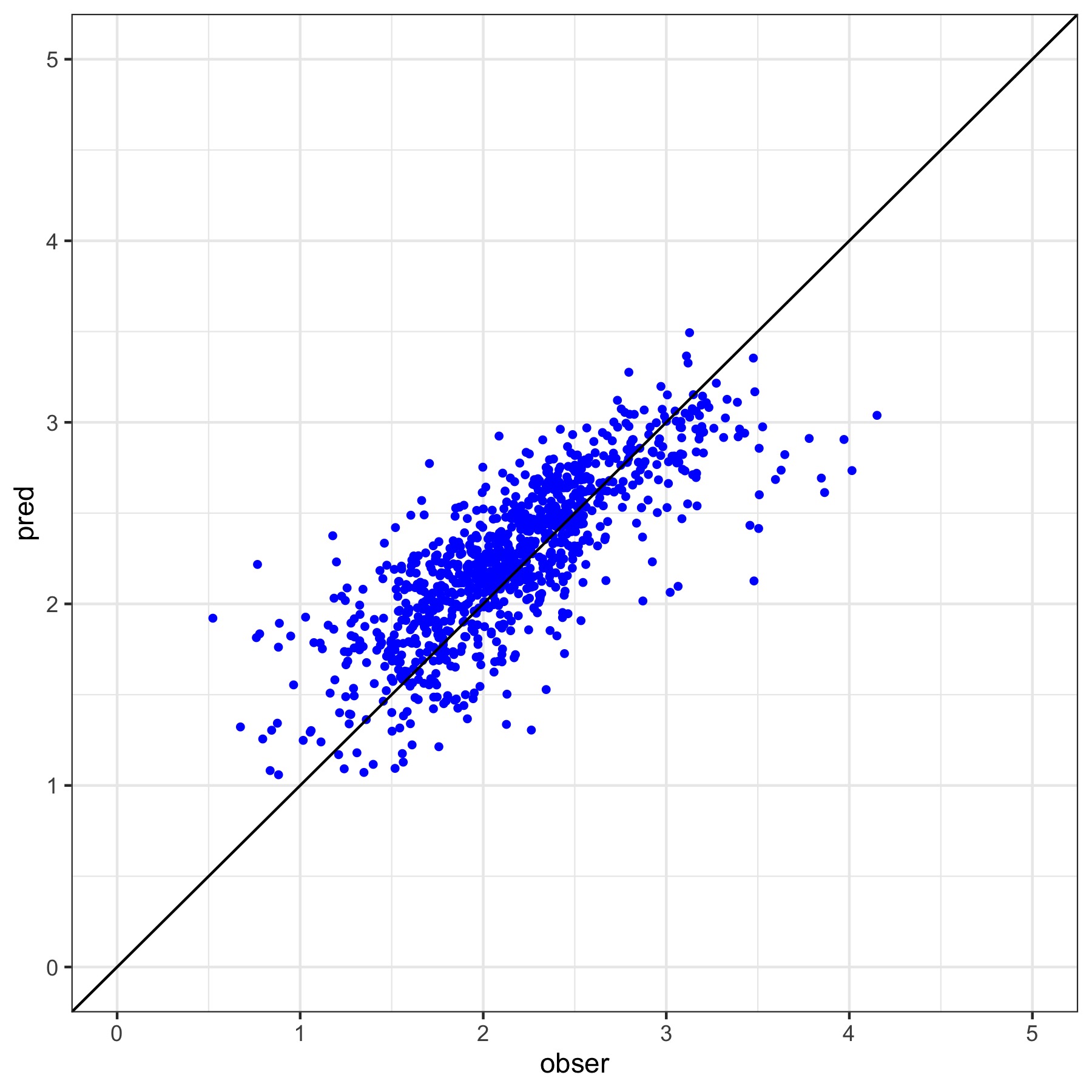}}
	\subfloat[]{
		\includegraphics[scale=0.1]{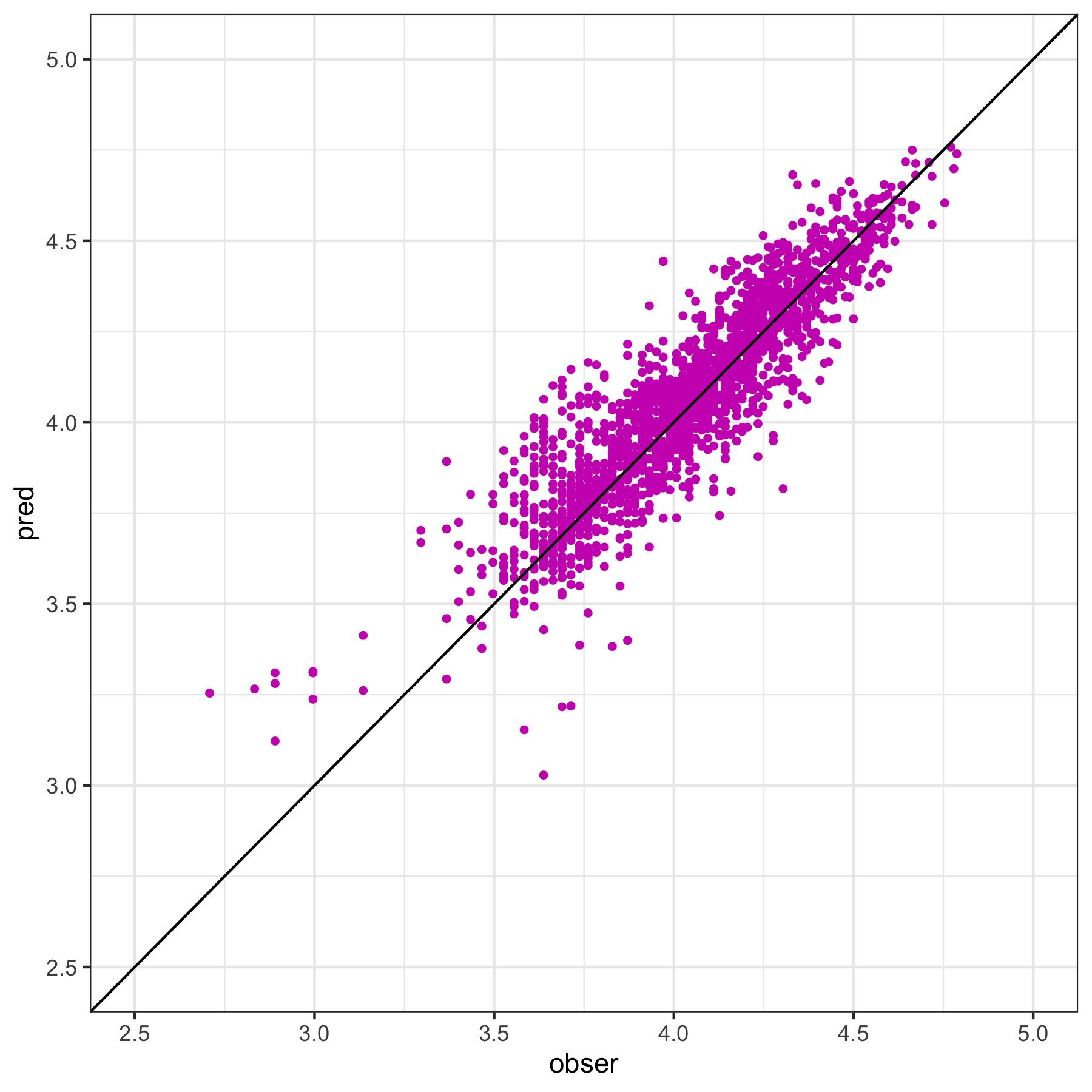}}\\  
  \caption{Visualisation of model prediction capability \CL{for PM$_{2.5}$ in (a,c)} 
  and ozone in (b,d). 
  Here (a,b) are for training sets in 2017--2020, and (c,d)  for testing sets in 2021. Model with points distributed along the identity line 
  \CL{indicates} the better one.}
	\label{obser_predict} 
\end{figure}

Fig.~\ref{obser_predict} shows the performance of Model 3 concerning its accuracy of model estimates (on the training set) and predictions (on the validation set). Compared with the PM$_{2.5}$ sub-figures, the scatters in O$_{3}$ sub-figures are clustered more closely around the identity line, indicating a stronger alignment between observed and estimated values. The PM$_{2.5}$
sub-model successfully captures overall trends but spreads out at the right tail (large values), aligning with the larger RMSE compared to the O$_{3}$ model.

\begin{table}[htbp]
\small 
   \centering
\caption{Model comparison for performance on the training and validation sets with different random effects specified in Models 1 $\sim$ 3. A higher correlation and lower values in all other criteria indicate better fitness.}
\resizebox{\textwidth}{!}{
\begin{tabular}{crrcccccrc}
\toprule
   \multirow{2}{*}{\begin{tabular}[c]{@{}c}Model \end{tabular}}     &   \multicolumn{3}{c}{Training Set (2017-2020)}   &\multicolumn{2}{c}{PM$_{2.5}$ Validation Set (2021)}  &\multicolumn{3}{c}{O$_{3}$ Validation Set (2021)}\\
          \cmidrule(r){2-4} \cmidrule(r){5-6}  \cmidrule(r){7-9}
       &  DIC & WAIC & LS & Correlation  & RMSE & Correlation   & \cl{RMSE} & \cl{StwCRPS ($p=0.7$)}  \\
\midrule
Model 1 & 43.33 & 3045.74  &  135169.72  & 0.39  & 0.52 & 0.74  & 0.20 & $0.35$\\
Model 2 & $-$4179.99 & $-$3003.04  &  119478.76  & 0.54   & 0.46 & 0.87 & 0.14 & $3.12$ \\
Model 3   &$-$9863.06 & $-$8608.15 &  102963.23  &  0.78  & 0.34 & 0.90 & 0.13 & 6.52 \\
\bottomrule
\end{tabular}
} \label{DIC}
\end{table}

\subsection{Summary and prediction}\label{model_visualization}
\begin{figure*} [htbp]
\centering
		\includegraphics[scale=0.15]{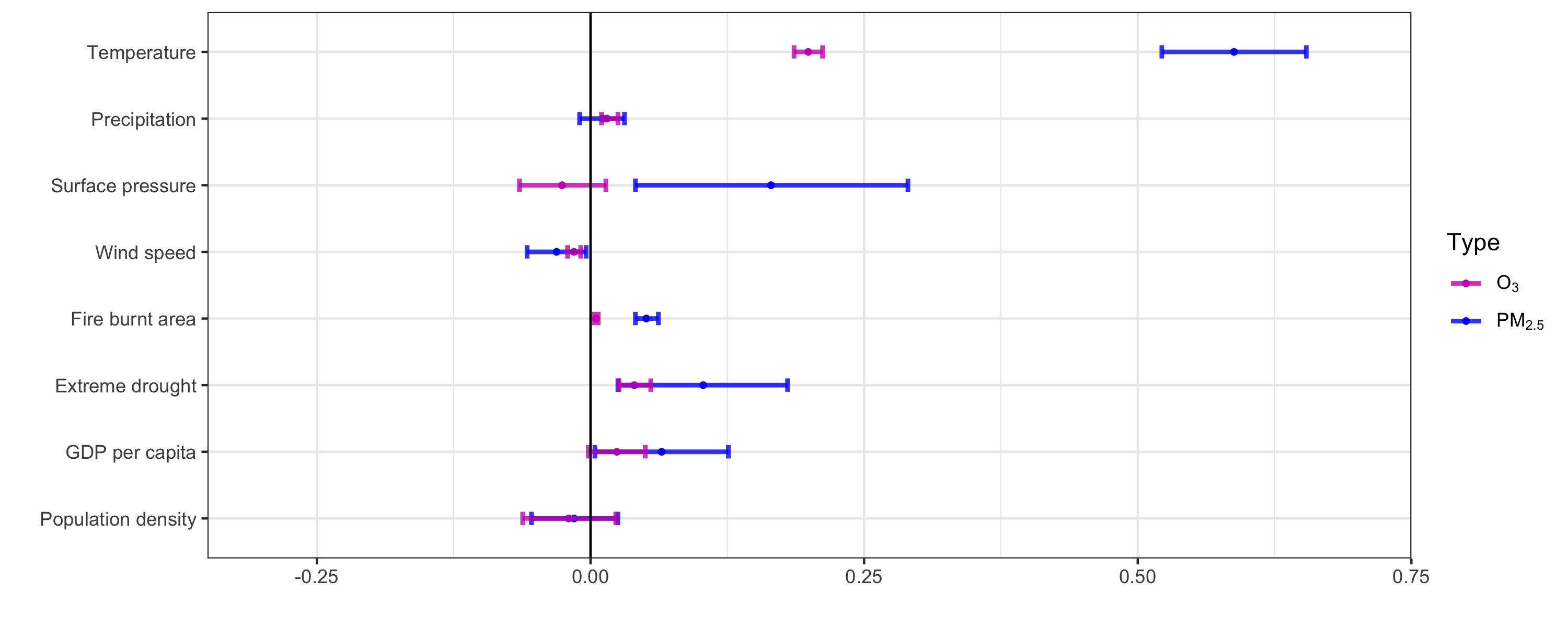}
\caption{Fixed effect quantiles plot. The three nodes in the quantiles plot indicate 0.025 quantile, mean and 0.975 quantile of the posterior estimates of each fixed effect associated with all predictors in Model 3.}
	\label{fixed_plot} 
\end{figure*}

Based on the optimal model (Model 3 with SPDE-AR(1) random effect), we summarized the posterior mean of fixed effects associated with all predictors (Table~\ref{tab: predictors}) together with its 95\% posterior credible interval (CI) in Table~\ref{pst_est_coefficient} with visualization in Fig.~\ref{fixed_plot}. The common drivers of PM$_{2.5}$ and ozone were identified, including temperature (positive association), extreme drought (positive), fire burnt area (positive), wind speed (negative association), and precipitation (both surface pressure and GDP per capita) shows only significant influence on ozone (PM$_{2.5}$), while population density affects neither. This observation suggests that human activities might not be as dominant in influencing air pollution compared to environmental processes \citep{WANG2020}.
Specifically, with one unit standard deviation ($\SI{23.16}{^\circ C}$) increase of temperature, monthly average PM$_{2.5}$ and maximum O$_{3}$ are expected 
to increase by 80\% (95\% CI: 68\%, 92\%) and 22\% (95\% CI: 20\%, 24\%), respectively. 
In addition, PM$_{2.5}$ and ozone in extreme drought month are expected to increase 
by a factor of 1.11 (95\% CI: 1.03, 1.20) 
and 1.041 (95\% CI: 1.026, 1.056), respectively. 
Meanwhile, 
the station located within a severe fire burnt area tends to have severe air pollutants, with 1.05 (\CL{95\% CI:} 1.04, 1.06) in PM$_{2.5}$ and 
1.005 (95\% CI: 1.003, 1.007) in O$_{3}$ times inclines with every 0.56 $\rm{km^2}$ increase of monthly cumulative burnt area, which suggests a direct connection between fire severity and air pollution. All the positive findings are consistent with the analysis of heatwaves in \cite{schnell2017co}, drought in  \cite{wang2017adverse} and wildfire in \cite{kalashnikov2022increasing}. 
In contrast, wind speed shows a significantly negative relationship with a decline by 3\% (95\% CI: 0.4\%, 5.6\%) in mean PM$_{2.5}$ and 1.48\% (95\% CI: 0.89\%, 2.08\%) in max O$_{3}$.
for every $11.4$m/s increase of monthly averaged wind speed, coinciding with the fact that the higher concentration is caused by stagnant winds limiting the horizontal dispersion of pollutants \citep{kinney2018interactions}. 
Finally, it is noted that 
surface pressure and precipitation are only positively associated 
with  PM$_{2.5}$ and 
ozone, respectively. Our result of positive association between precipitation-O$_3$ is in consistency with that found by 
\cite{Arshinova2019}. There are still some other studies showing the inverse (negative) impacts of precipitation on ozone \citep{CAO2020,LIU2019}.

\begin{table}[!htbp] \centering
 \tiny
\caption{Posterior mean (95\% credible interval) of the fixed effects.}
\setlength{\tabcolsep}{4.5mm}
\resizebox{\linewidth}{!}{  
\begin{tabular}{*{5}{lrcrc}}
\toprule
   \multirow{2}*{Covariate} & \multicolumn{2}{c}{Month mean model ($\boldsymbol{\beta}_{\text{PM}}$)} & \multicolumn{2}{c}{Month maxima model ($\boldsymbol{\beta}_{\text{OZ}}$)}  \\
  \cmidrule(rr){2-3}\cmidrule(lr){4-5}
  & Mean & 95\% CI & Mean &  95\% CI  \\
  \midrule
  {Temperature} & $0.588$ &$ (0.522,  0.654)$ & $0.199$ &$ (0.186, 0.212)$ \\
  {Precipitation} & $0.011$ & $ (-0.010, 0.031)$ & $ 0.021$ &$ (0.017, 0.025)$ \\  
  {Surface pressure} & $0.165$ & $ (0.041, 0.290)$ & $-0.026$ & $ (-0.065, 0.014)$  \\
  {Wind speed} & $-0.031$ & $ (-0.058, -0.004)$ & $-0.015$ & $ (-0.021, -0.009)$  \\
  {Fire burnt area} & $0.051$ & $ (0.041, 0.062)$ & $0.005$ & $ (0.003, 0.007)$ \\
  {Extreme drought} & $0.103$& $ (0.025, 0.180)$ & $0.040$ &$(0.026, 0.055)$ \\
  {GDP per capita} & $0.065$& $ (0.004, 0.126)$ & $0.024$ &$ (-0.002, 0.050)$ \\
  {Population density } & $-0.015$& $ (-0.054, 0.025)$ & $-0.020$ &$ (-0.062, 0.023)$ \\
  \bottomrule
\end{tabular}}    
\label{pst_est_coefficient}
\end{table}

The posterior means, standard deviations, and 95\% credible intervals of parameters are presented in Table~\ref{pst_est_random}. The posterior mean of the range parameter $\rho_s$ is 2.92  (95\% CI: 2.56, 3.37) in the sharing SPDE-AR(1) model and $\rho'_s$ is 2.85 (95\% CI: 2.46, 3.28) in the additional model. Both range parameters are significantly below 10, which is aligned with the PC prior and implies that spatial dependence is even weaker. The auto-correlation coefficients $a_s,\ a'_s,\ a_y$ are estimated as 0.91, 0.98 and 0.56, respectively. This indicates strong 
\CL{inner-annual and inter-annual dependence},
aligning with the temporal trends presented in the box-plot (Fig.~\ref{boxplot}).   More importantly,  \CL{the coefficient} ($\beta_5$) for sharing SPDE-AR(1) term is significantly negative with value \CL{$-0.708$ (95\% CI: $-0.896, - 0.476$),} 
suggesting an \CL{inverse sharing of the spatio-temporal effect between moderate PM$_{2.5}$ and extreme O$_3$ processes}. 
This evidence coincides with the findings of the enhanced PM$_{2.5}$ levels suppress surface solar radiation and could weaken O$_{3}$ production with the low diurnal peaks \citep{Jia2017}.

\begin{table}[ht]
    \centering
\small    
\caption{Posterior estimates of mean, standard deviation and quantiles of the parameters. The precision of Gaussian ($\tau_{\text{PM}}$) and Gumbel distributions ($\tau_{\text{OZ}}$), the range parameter ($\rho_{s}$), the standard deviation ($\sigma_{s}$) and the autocorrelation coefficients ($a_{s}$) in the sharing SPDE-AR(1) model, $(\rho'_{s},\sigma'_{s},a'_{s})$ in the extra SPDE-AR(1) model, precision ($\tau_{y}$) and autocorrelation coefficient ($a_{y}$) in the yearly AR(1) model, precision of measurement errors ($\tau_{1},\tau_{2}$), and sharing effect coefficients ($\beta_4, \beta_5$).}
\setlength{\tabcolsep}{0.15mm}
\begin{tabular*}{\linewidth}{@{\extracolsep{\fill}}lrrrrr}
    \toprule  Parameter  & Mean & Stdev & 0.025 quantile & 0.5 quantile & 0.975 quantile \\
    \midrule $\tau_{\text{PM}}$  & 8.466  & 0.185 & 8.081 & 8.470  & 8.817 \\
           $\tau_{\text{OZ}}$  & 212.865 & 7.673 & 197.160 & 213.067 & 227.423 \\
           $\rho_{s}$  & $2.917$ & 0.204 & $2.557$ & $2.903 $ & $3.370$ \\
           $\sigma_{s}$ & 0.618 & 0.030 & 0.565 & 0.616  & 0.685\\
           $a_{s}$ &0.907 & 0.009 & 0.891 & 0.907 & 0.925 \\
           $\rho'_{s}$  & $2.851$ & 0.202 & $2.457$ & $2.845 $ & $3.283$ \\
           $\sigma'_{s}$ & 0.461 & 0.029 & 0.405 & 0.460  & 0.522\\
           $a'_{s}$ &0.976 & 0.002 & 0.971 & 0.976 & 0.981 \\
           $\tau_{y}$ & 43.959 & 7.083 & 31.112 & 43.575 & 59.077 \\
           $a_{y}$ & 0.561 & 0.039 & 0.476 & 0.563 & 0.631 \\
           $\tau_{1}$ & 1276.212 & 108.649 & 1097.029 & 1266.046 & 1528.065\\
           $\tau_{2}$ & 309.929 & 16.141 & 281.197 & 308.852 & 344.901\\
           $\beta_4$ & 0.005 & 0.017 & $-$0.027 & 0.005 & 0.039 \\
           $\beta_5$ &  $-$0.708 & 0.107 & $-$0.898  & $-$0.714 & $-$0.476 \\\bottomrule
    \end{tabular*}
    \label{pst_est_random}
\end{table}
\begin{figure*} [htbp]
\centering
  \subfloat[]{
		\includegraphics[scale=0.08]{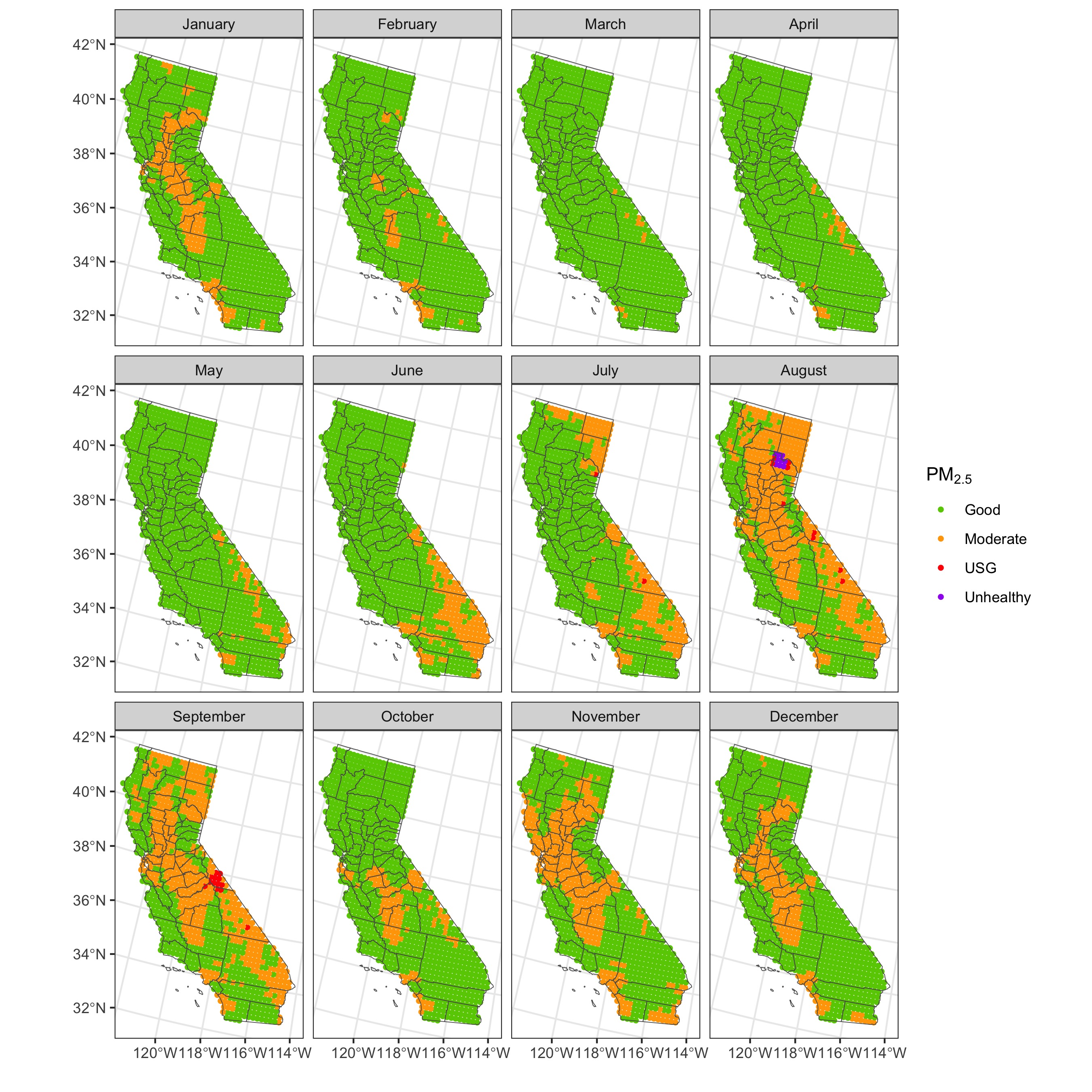}}
  \subfloat[]{
		\includegraphics[scale=0.08]{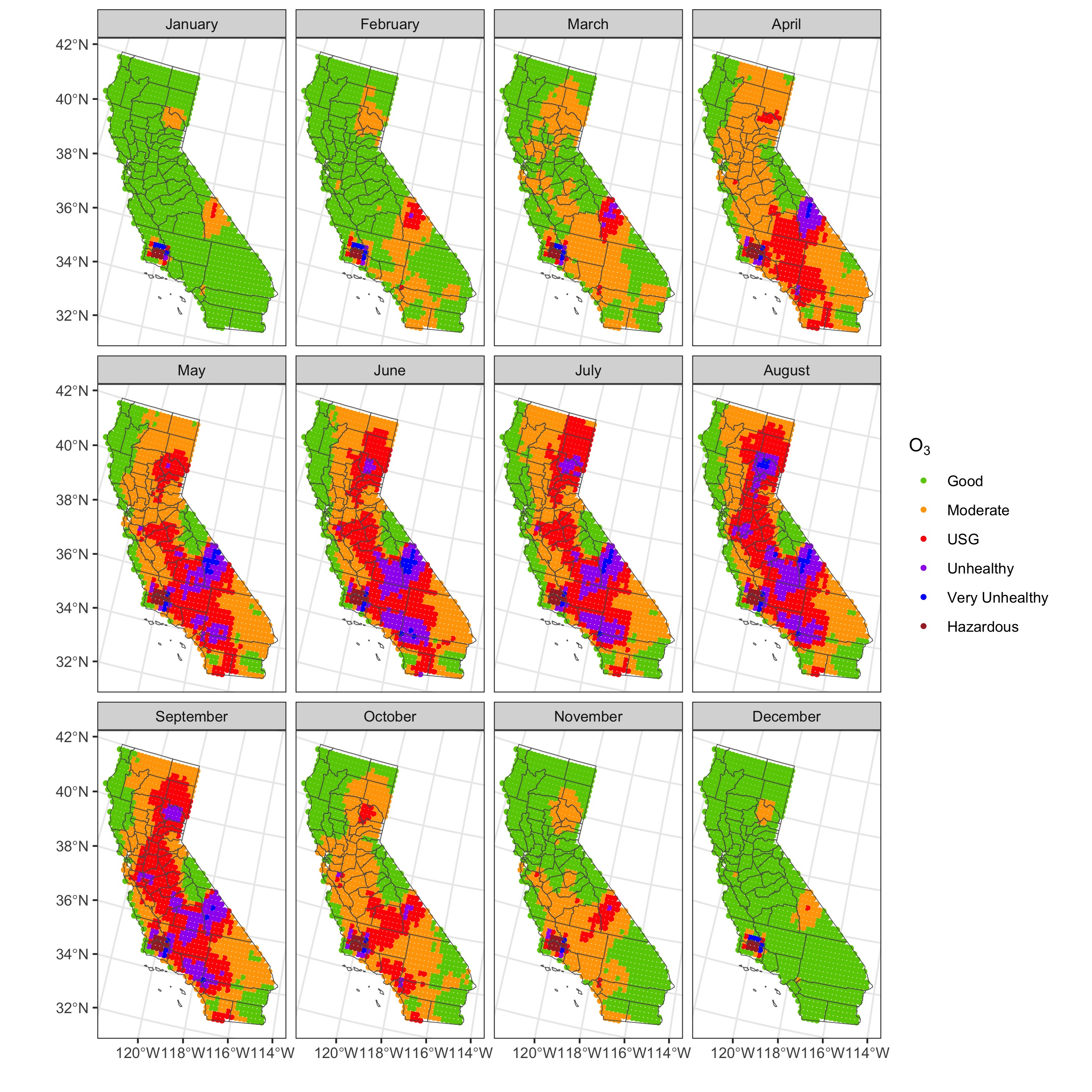}}
\caption{Prediction of monthly mean PM$_{2.5}$ (a) and monthly max ozone (b) over the entire California from January to December, 2021 based on Model 3.
} \label{full_pred}
\end{figure*}
\CL{Based on Model 3 in the training set (2017--2020), Fig.~\ref{full_pred} presents the prediction of PM$_{2.5}$ and ozone across the entire state of California for each month in 2021.}
Generally, the concerns regarding PM$_{2.5}$ are more noticeable in August, September, November and December. Conversely, monthly maximums of ozone peak in hot seasons, specifically from April to September. These inverse temporal variation patterns of air pollutants align with our  empirical analysis 
(Fig.~\ref{boxplot}). Regarding spatial distribution, northern California appears to be more susceptible to experiencing moderate to unhealthy levels of PM$_{2.5}$, while southern California is more likely to experience extreme levels of ozone pollution sustainably. It is noteworthy that the area around the intersection of San Luis Obispo and Santa Barbara counties consistently exhibit maximum ozone levels exceeding the hazardous threshold throughout the year.

In air pollution research, of key is to identify regions with high pollution levels and predict air pollution patterns. These methods help visualize air quality conditions for the public and environmental regulators. 
The positive $\mathrm{E}_{u, \alpha}^{+}(X)$  and negative $\mathrm{E}_{u, \alpha}^{-}(X)$ excursion sets, proposed by \cite{bolin2015excursion}, determine the largest sets that simultaneously exceed or fall below the risk level $u$ with a small error probability  $\alpha$, utilizing a parametric family in conjunction with a sequential importance sampling technique to estimate joint probability distributions. To visualize the excursion sets simultaneously, we apply the positive and negative excursion functions,  $F_{u}^{+}(s)=1-\inf \left\{\alpha \mid s \in \mathrm{E}_{u, \alpha}^{+}\right\}$ and  $F_{u}^{-}(s)=1-\inf \left\{\alpha \mid s \in \mathrm{E}_{u, \alpha}^{-}\right\}$. The term  $\inf \left\{\alpha \mid s_{0} \in \mathrm{E}_{u, \alpha}^{+}\right\}$ denotes the "smallest"  $\alpha$  required for the location  $s_{0}$ to be included into the positive excursion set  $\mathrm{E}_{u, \alpha}^{+}$, 
while the higher  $1-\inf \left\{\alpha \mid s_{0} \in \mathrm{E}_{u, \alpha}^{+}\right\}$ reported by positive excursion function generally indicates higher \CL{confidence/likelihood} 
for the location  $s_{0}$ to exceed the risk threshold $u$ simultaneously. 

\begin{figure*} [htbp]
\centering
	\subfloat[\label{fig7:a}]{
		\includegraphics[scale=0.08]{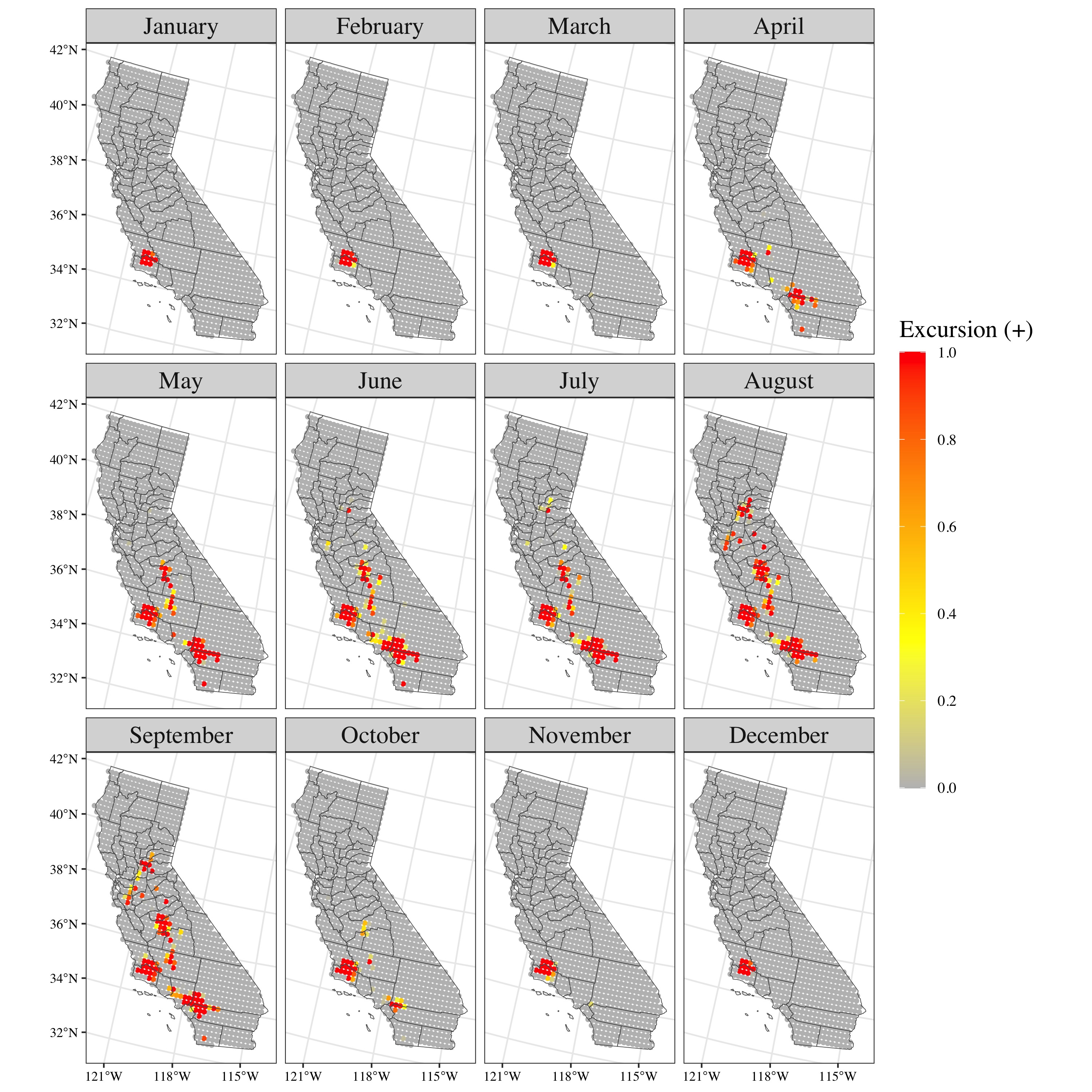}}
	\subfloat[\label{fig7:b}]{
		\includegraphics[scale=0.08]{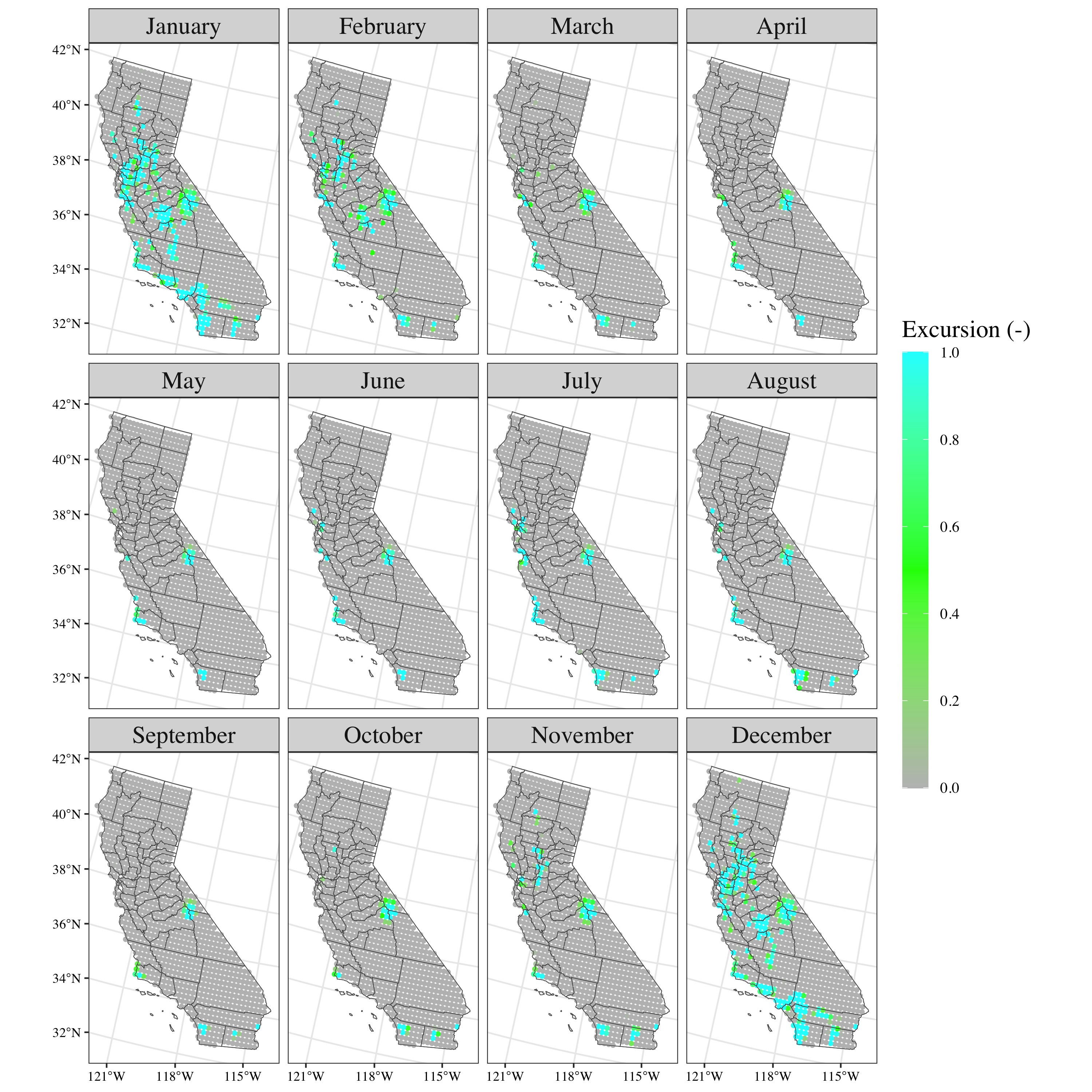}}
\caption{\CL{Positive and negative  excursion functions  of monthly maximum ozone from January to December in 2021, respectively. Here the thresholds for positive excursion function in (a) and negative excursion function in (b) are respectively 71 ppb and 54 ppb,  indicating the "Unhealthy for sensitive groups" and "Good" categories of ozone according to the EPA in the U.S.}
}
	\label{excursion} 
\end{figure*}

We set $\alpha=0.05$ and present in Fig.~\ref{excursion}  for the positive excursion function with \CL{$u= 71$ ppb} and the negative excursion function with \CL{$u= 54$ ppb}. The two risk levels denote \CL{the cut-offs for} 'Unhealthy for sensitive groups' and \CL{'Good'} categories for ozone concentrations, respectively. 
We see from Fig.~\ref{excursion}(a) that, there is a consistently high probability for 
a set of areas with unhealthy ozone levels for sensitive populations, including 
San Luis Obispo and Santa Barbara counties. Additionally, 
the middle part of California are also prone to unhealthy ozone concentrations from May to October.
Fig.~\ref{excursion}(b) presents the areas considered relatively 'safe' concerning poor ozone air quality. More areas in winter 
maintain healthy O$_3$ quality simultaneously than in the summer. 
Areas around the intersection of Mono, Fresno and Inyo  
are likely to display good O$_{3}$ quality  aligning with other parts of the state 
throughout the year.

\section{Discussion and conclusion} \label{section_discussion}
We proposed a joint spatio-temporal framework allowing for sharing stochastic interactions between extreme O$_3$ and moderate PM$_{2.5}$. While the moderate PM$_{2.5}$ concentrations are (marginally) modelled using a log-Gaussian random field, the extreme O$_3$ levels are well fitted by the log-Gumbel model from extreme value theory. The high flexibility of our new models suggests that they can be more generally utilized to jointly model multiple processes in other contexts, e.g., extreme climate and environmental events as well as multiple interrelated diseases.

Our joint spatio-temporal Bayesian hierarchical model identified common and different drivers of the cumulative effect of PM$_{2.5}$ and the extreme effect of O$_3$ in California, providing new insights into the direction and magnitude of common drivers. The common drivers of temperature, extreme drought, fire burnt area and wind speed display significant influence on extreme O$_3$ with less uncertainty than that on PM$_{2.5}$. The corresponding findings are consistent with those identified by \cite{schnell2017co, wang2017adverse, kalashnikov2022increasing}. As suggested by \cite{chen2020understanding}, the inner causal effect of wind (i.e., wind directions together with speeds) and air pollutants can be further investigated. To conclude, our result demonstrated the improved inferences through sharing in our air pollution modelling.  Similar advantages of joint sharing modelling are illustrated in the study of air pollution \citep{wang2023spatio}, wildfire \citep{koh2023spatiotemporal, zhang2023joint}, landslides \citep{yadav2023joint} and disease \citep{alahmadi2024joint}.  

The joint PM$_{2.5}$-O$_3$ model provides new insights into  \cling{appropriate} risk controls and region-specified strategies. 
\begin{itemize}
\item  First, the reverse spatio-temporal interaction is demonstrated with a significant negative scaling coefficient, 
This discovery 
highlights the potential distinct spatio-temporal patterns of PM$_{2.5}$ and ozone in California. \CL{Similar findings were identified by \cite{Jia2017}  in East China. In addition, the peaks of PM$_{2.5}$ and ozone appear respectively in cold and hot seasons,} 
which is in correspondence with the observed inverse association between these two pollutants. Specifically, \CL{as shown by  \cite{Jia2017}, elevated O$_3$ levels in hot season might} 
stimulate the formation of secondary particles. In contrast, during the cold season, PM$_{2.5}$ is likely to suppress surface solar radiation, thereby weaken O$_3$ production. 
\item Second,  the Bayesian framework of our joint {continuous-time spatio-temporal models 
}
enables us to predict air pollutant conditions  individually and identity hot-spots using excursion functions (for ozone)   throughout the whole California. Our results suggest more attention to controlling serious PM$_{2.5}$ and O$_3$ pollution in the middle and north of California, respectively. 
Meanwhile, relevant agencies need to take seasonal actions when it comes to higher PM$_{2.5}$ concentration in both autumn and winter and O$_3$ in summer. 
Further, in the aid of excursion functions, the identified area around intersection of San Luis Obispo and Santa Barbara counties is likely to exceed simultaneously the unhealthy O$_3$ level  for sensitive population throughout the year. These findings not only enhance our understanding of air pollution dynamics but also serve as valuable insights for future policy making and resources allocation regarding the prevention and treatment of air pollution.
\end{itemize}

To conclude, we have implemented a novel joint Bayesian spatio-temporal model for multiple air pollutants, with shared random effects to account for stochastic dependence among different model components not explained by covariates. Different sharing strategies should be incorporated for different considerations \citep{koh2023spatiotemporal}. In this study, we emphasized on accurate inference of extreme ozone for its large spatial and temporal disparities. Our findings improve decision support in the co-control of PM$_{2.5}$ and ozone at spatial scale.  Future work could better explore  both linear and non-linear effects of covariates by stochastic splines \citep{alahmadi2024joint}.

\bibliography{references.bib}{}
\bibliographystyle{apalike}


\end{document}